\documentclass[showpacs,11pt]{revtex4}

\usepackage{epsfig}
\usepackage{amssymb}

\newcommand{\figurewidth}{3in}

\newcommand\average[1]{\langle#1\rangle}

\newcommand\eqns[2]{(\ref{#1})-(\ref{#2})}
\newcommand\fig[1]{Fig.~\ref{#1}}
\newcommand\figs[2]{Figs.~\ref{#1}-\ref{#2}}

\begin{document}


\title{Compact stars in the standard model - and beyond}

\author{F. Sandin\footnote{e-mail: \tt fredrik.sandin@ltu.se}}

\affiliation{Department of Physics, Lule{\aa} University of Technology, SE-97187 Lule\aa , Sweden}

\begin{abstract}
In the context of the standard model of particle physics, there is a definite upper limit to the density of
stable compact stars. However, if there is a deeper layer of constituents, below that of quarks and leptons,
stability may be re-established far beyond this limiting density and a new class of compact stars could exist.
These objects would cause gravitational lensing of white dwarfs and gamma-ray bursts, which might be observable
as a diffraction pattern in the spectrum. Such observations could provide means for obtaining new clues about
the fundamental particles and the origin of cold dark matter.
\end{abstract}


\maketitle


\section{Introduction}

The different types of compact objects traditionally considered in astrophysics are white dwarfs, neutron stars
(including quark and hybrid stars), and black holes. The first two classes are supported by Fermi degeneracy
pressure from their constituent particles. For white dwarfs, electrons provide the pressure counterbalancing
gravity. In neutron stars, the neutrons play this role. For black holes, the degeneracy pressure is overcome
by gravity and the object collapses to a singularity, or at least to the Planck scale
($\rho \sim 10^{93}$~g/cm$^3$). For a recent review of neutron stars, hybrid stars, and quark stars, see,
{\it e.g.},~\cite{Weber:04} and references therein.

The distinct classes of degenerate compact stars originate directly from the properties of gravity, as
was made clear by a theorem of Wheeler and collaborators in the mid 1960s~\cite{Harrison:65}.
This theorem states that for the solutions to the stellar structure equations, whether Newtonian or
relativistic, there is a change in stability of one radial mode of vibration whenever the mass reaches
a local maximum or minimum as a function of the central density. The theorem assures that distinct
classes of stars, such as white dwarfs and neutron stars, are separated in central density by a region
in which there are no stable configurations.

In the standard model of quarks and leptons (SM), the theory of the strong interaction between quarks
and gluons predicts that with increasing energy and density, the coupling between quarks asymptotically
fades away~\cite{Gross:73,Politzer:73}. As a consequence of this so-called asymptotic freedom, matter is
expected to behave as a gas of free fermions at sufficiently high densities. This puts a definite upper limit
to the density of stable compact stars, since the solutions to the stellar equations end up in a never-ending
sequence of unstable configurations, with increasing central density. Thus, in the light of the standard
model, the densest stars likely to exist are neutron stars, quark stars, or the slightly more dense hybrid
stars~\cite{Gerlach:68,Glendenning:00,Schertler:00}. However, if there is a deeper layer of constituents,
below that of quarks and leptons, the possibility of a new class of compact stars opens up~\cite{Hansson:04}.

Though being a quantitatively successful theory, the SM consists of a large number of exogenous {\it ad hoc}
rules and parameters, which were introduced solely to fit the experimental data. The SM provides no explanation
for the deeper meaning of these rules. At a closer look, however, the SM seems to be full of hints to its deeper
background. By considering these rules from a historical point of view, a ``simple" and appealing explanation
is {\it compositeness}~\cite{Fredriksson:03}, {\it i.e.}, that the quarks, leptons, and gauge bosons are composite
particles, built out of more elementary {\it preons}~\cite{dSouza:92}. Preons provide natural explanations for
the particle families of the SM and phenomena such as neutrino oscillations, mixing of the weak gauge bosons,
and quarks of different flavour.

Over the last decades, many papers have been written about preons, but so far there are no direct
evidence for (or against!) their existence. In the late 1970s, a number of consistency conditions were
formulated by 't~Hooft~\cite{tHooft:79}. In the same work, a vector-like non-Abelian $SU(3)$ gauge group was
considered, but no solution to the consistency conditions was found. Later, it was shown that with another
choice for the gauge group and the flavour structure of preons, {\it e.g.}, three different preon flavours,
the consistency conditions are satisfied~\cite{Barbieri:80}. For a more detailed discussion of preon models,
see~\cite{Fredriksson:03,dSouza:92,PreonTrinity:02} and references therein.

Not all clues favour preon models, but the existence of preons is still an open question, and as a consequence
so is the question whether a new class of compact stars exists or not. This paper is based on the ideas and
results presented in~\cite{Hansson:04}. Assuming that quarks and leptons are composite particles, built out of
more elementary preons, I will:
\newcounter{Lcount}
\begin{list}{\Roman{Lcount}.}{\usecounter{Lcount}}
\item{Give an estimate for the mass and radius of stars composed of preons.}
\item{Show that for some particular equations of state, stable solutions to the general relativistic
	stellar equations do exist, with densities far beyond the maximum density in stars composed of
	quarks and leptons.}
\item{Briefly discuss some potential astrophysical consequences and how these objects could be
	observed. Herein lies the potential importance of this qualitative speculation, since these
	objects are candidates for cold dark matter and could be found as, {\it e.g.}, gravitational
	``femtolenses".}
\end{list}


\section{Compact stars and the maximum density prophecy}

In order to explain why there is a maximum density for stars composed of quarks and leptons, or any other
composite particle composed of these two species, {\it e.g.}, nucleons and $^{56}$Fe, some basic knowledge
about the theory of compact stars is needed. In the following, I give a short introduction and a
summary of the main points.

Due to the high density and large mass of compact stars, a general relativistic treatment of the equilibrium
configurations is necessary. This is especially important for the analysis of stability when a star is
subject to radial oscillations. Such oscillations are excited to some extent, and for a star to be stable
the amplitude of the oscillations must not grow spontaneously with time. The starting point for a general
relativistic consideration of compact stars is the Oppenheimer-Volkoff (OV) equations~\cite{OV} for
hydrostatic, spherically symmetric equilibrium:
\begin{eqnarray}
\frac{dp}{dr} &=& -\frac{G\left(p+\rho c^2\right)\left(m c^2 + 4 \pi r^3 p\right)}
	{r \left(r c^4 - 2 G m c^2 \right)}, \label{ov1} \\
\frac{dm}{dr} &=& 4 \pi r^2 \rho. \label{ov2}
\end{eqnarray}
Here $p$ is the pressure, $\rho$ the density and $m=m(r)$ the mass within
the radial coordinate $r$. The total mass of the star is:
\begin{equation}
M = 4\pi \int_0^R r^2 \rho\,dr,
\end{equation}
where $R$ is the coordinate radius of the star. Combined with an equation of state (EOS) $p = p(\rho)$
obtained from some microscopic (quantum field) theory, the OV solutions give the possible equilibrium
states of spherically symmetric stars.

\begin{figure}[ht]
\epsfig{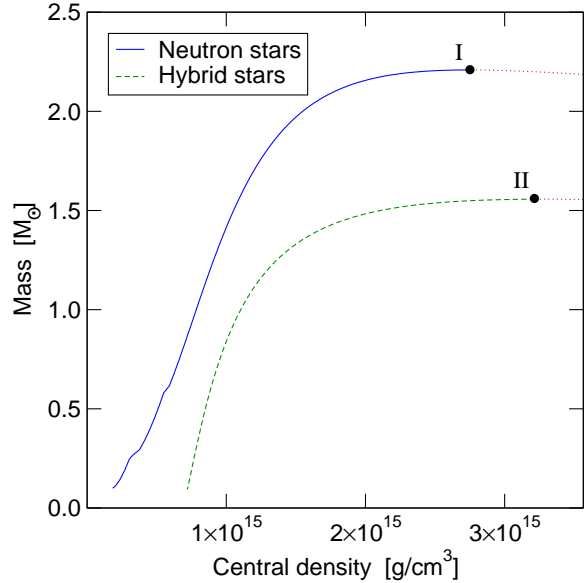}
\caption{Two sequences of compact stars obtained by solving the OV equations. The hybrid stars have a
	core of unpaired quarks and the nuclear matter crust extends down to $1$\% of the nuclear saturation density.
	The neutron star sequence is stable up to the maximum mass configuration (I).
	For this particular equation of state, this configuration has the highest possible (central) density,
	as stars more massive and dense than this are unstable and collapse into black holes. The stable hybrid
	star sequence terminates at II. M$_\odot\simeq2\times10^{30}$~kg is the solar mass.}
\label{ordinary_stars}
\end{figure}

As an example, I show two sequences of compact star configurations. One is composed of nuclear matter
(neutron stars) and the other of a deconfined quark matter core with a nuclear matter crust
(hybrid stars), see~\fig{ordinary_stars}.
%
%
These configurations were obtained by solving the OV equations~\eqns{ov1}{ov2} numerically. The low-density
part of the nuclear matter EOS was extracted from~\cite{NV:71} and the high-density part comes
from~\cite{APR:98}. For the deconfined quark matter phase an unpaired massless quark approximation
$\rho c^2 = 3p + 4B$ was used, and the ``bag pressure" $B$ was fitted such that the transition from quark
matter to nuclear matter occurs at $0.01$ times the nuclear saturation density, $n_0 \sim 0.16$~fm$^{-3}$.
The exact density where cold nuclear matter decompose into quark matter is unknown, so the transition density
used here serves as an example only.

The composition of matter at neutron star densities is an open question and many different models for the
EOS exist, {\it e.g.}, EOSs for nuclear matter, matter with hyperons, and superconducting quark matter.
Regardless of the specific model, the maximum mass and corresponding radius are roughly one solar mass
$M_\odot \simeq 2\times 10^{30}$~kg
and $10$~km. No substantially more dense configurations composed of quarks and leptons are possible. The motivation
goes roughly like this: At white dwarf densities, the nucleons occupy nuclei that contribute little to the pressure,
and electrons provide the pressure counterbalancing gravity. With increasing density, the pressure rises and the
electrons become more energetic. Eventually, the electrons are captured by protons and the pressure drops. As a
consequence, the white dwarf sequence becomes unstable and terminates. At roughly six to seven orders of magnitude
higher density than in the maximum-mass white dwarf, nuclei dissolve and the Fermi pressure of nucleons (in nuclear
matter) and quarks (in quark matter) stabilize the next sequence of stable stars. The maximum mass of this sequence
is of the order of one solar mass, for all compositions (nuclear matter, strange matter, hyperons etc.).

The reason why this is the limiting mass of stable compact stars composed of quarks and leptons is simply that
there is no particle that may stabilize another sequence of stars. Each quark flavour is accompanied by an extra
Fermi sea that relieves the growth of pressure, and quark Fermi pressure is only won at the expense of pressure
from other species. Also, the chemical potential in stable hybrid stars is much smaller than the charm mass, so
stars composed of quarks heavier than the strange quark do not exist~\cite{Kettner:95,Prisznyak:94}.

Hence, beyond the very rich and beautiful landscape of structures composed of quarks and leptons, 
at $\sim 10^{16}$~g/cm$^3$, there is again a desert of instability, just like there are no stable stellar
configurations in-between white dwarfs and neutron stars. The question is now if the desert ends before
the Planck scale.


\section{Compact stars beyond the desert}

A definite upper limit to the density of any static spherically symmetric star can be obtained from the
Schwarzschild radius,
\begin{equation}
R = 2GM/c^2,
\label{schwarz}
\end{equation}
since any object more dense than this would collapse into a black hole. By using the expression for the
Schwarzschild radius and the relations:
\begin{eqnarray}
M &\sim& m A, \label{msimple} \\
R &\sim& d_0 A^{1/3}, \label{rsimple}
\end{eqnarray}
where A is the number of constituent particles, $m$ their mass and $d_0$ the distance
between adjacent particles, an order of magnitude estimate for the maximum mass and radius of the corresponding
class of compact stars can be calculated. For an idealized neutron star composed of nucleons of mass
$m_n \simeq 939$~MeV$/c^2$ and size $d_n\simeq 0.5\times 10^{-15}$~m~\eqns{schwarz}{rsimple} give
$A\sim 3\times 10^{57}$~baryons, $R\sim 7$~km and $M \sim 5 \times 10^{30}$~kg$\,\sim 2.5$~M$_\odot$. In
reality, a somewhat larger radius and smaller mass are expected, since the density is non-uniform in the star,
say $R \sim 10$~km and $M \sim 2$~M$_\odot$. In any case, the correct order of magnitude for the maximum mass
and corresponding radius of a neutron star is obtained. The average density is $\bar{\rho}\simeq 10^{15}$~g/cm$^3$.

Since the Schwarzschild limit is almost reached already for the most massive neutron stars, it is reasonable
to assume that this should be the case also for a more dense class of compact stars. Then, in order to provide
similar estimates for the mass and radius of a star composed of preons, something must be known about the mass
and ``size" of preons. Before trying to achieve this, it should be emphasized that we know nothing about preons,
not even if they exist. So whatever method used, the result is a speculative order of magnitude estimate.
But as I will show, it is still possible to reach some qualitative results.

Guided by the observation that the density of nuclear matter is roughly of the same order of magnitude as for
deconfined quark matter, I assume that the density of preon matter is roughly of the same order of magnitude
as for a closely spaced lattice of some ``fundamental" particle of the SM. In this case the problem is simplified
to finding a fundamental SM particle, with known mass and maximum spatial extension.
The simplest and least ambiguous choice seems to be the electron, since the mass of an electron is well known,
and from scattering experiments it is known that electrons do not have any visible substructure down to a
scale of $\sim \hbar c/$TeV$\,\sim 10^{-19}$~m. Using the electron mass $m_e \simeq 511$~keV$/c^2$ and an
upper estimate for its radius $r_e \sim 10^{-19}$~m, the maximum mass and radius of a star composed of preons
is found to be of the order $M\sim 10^2$~M$_\oplus$ and $R\sim 1$~m. Here M$_\oplus\simeq 6\times10^{24}$~kg is
the mass of the Earth. The average density is of the order $\sim 10^{23}$~g/cm$^3$.

This crude estimate gives metre-sized objects that are a hundred times more massive than the Earth.
Now, I will try to be a bit more specific. Especially, it would be interesting to see whether such objects
could be stable or not. In order to do this, I extrapolate an effective model for hadrons, the
so-called MIT bag model~\cite{Chodos:74}. In its simplest form, the MIT bag is a gas of massless
fermions (partons), enclosed in a region (the bag) and subject to an external pressure $B$ (the bag constant).
The EOS for a gas of massless fermions is $\rho c^2=3 p$ and by including $B$ one obtains:
\begin{equation}
\rho c^2=3p+4B.
\end{equation}
This result does not depend on the degeneracy factor, {\it i.e.}, the number of fermion species, spin, etc. For
a single hadron the pressure is practically zero, so that $\rho c^2=4B$ and the total energy $E$ of a hadron
is~\cite{Chodos:74}:
\begin{equation}
E=4B\average{V},
\end{equation}
where $\average{V}$ is the time-averaged volume of the bag (hadron). Hence, the bag pressure $B$ must be of
the order of $1$~GeV/fm$^3\sim 10^{35}$~Pa for hadrons. This is in agreement with experiments and other
independent methods of calculating light-quark hadronic masses; Most of the mass-energy is not due to
the ``bare mass" of the constituents, but the confining interactions.

The MIT bag model is frequently used for the description of deconfined quark matter and applications to
compact stars. Its usefulness in this regime originates in asymptotic freedom, simplicity and the possibility
to include pertubative corrections. The bag pressure $B$ is introduced in order to confine partons, it is a
phenomenological parametrization of the strong interactions that confine quarks into hadrons. These interactions
are present also in deconfined quark matter, so the ``bag model" should be applicable also in this regime. However,
the value of the bag pressure is different, since the density is higher and the interactions weaker. Thus, the
so-called bag constant $B$ is not really a constant, but a density dependent parameter. For strange quark matter,
the bag constant is roughly $B^{1/4} \sim 150$~MeV$/(\hbar c)^{3/4}$~\cite{Glendenning:97} and the corresponding
contribution to the energy density is $4B \sim 260$~MeV/fm$^3$. This means that a considerable fraction of the
density in quark matter, roughly $10^{15}$~g/cm$^3$, is due to the bag constant, {\it i.e.}, interactions.

Now, the fundamental assumption here is that preons exist and are fermions. Since preons constitute light particles,
such as neutrinos and electrons, the ``bare" preon mass should be fairly small. Then a massless fermion approximation,
$\rho c^2=3p$, can be used. This EOS does not allow for stable super-dense stars, however, so something more is needed.
And that `something' is dynamics, the preon interactions that give mass-energy to the particles of the SM.
The question is how, since there is so far no quantitative model for preon interactions. Indeed, a fundamental problem
in preon models is to find a suitable dynamics, capable of binding preons into fermions with masses essentially negligible
with respect to their inverse radius. With this in mind, the principle of parsimony (``Occam's razor") seems to be the
only guidance. 

A simple solution is to include the dynamics in terms of a bag constant~\cite{Hansson:04}, which roughly reproduces
the minimum energy density of an electron,
\begin{eqnarray}
	B &=& \frac{E}{4\average{V}} \sim \frac{3\times 511\,{\rm keV}}{16\pi(10^{-19}\,{\rm m})^3}
	\sim 10^{4}\,{\rm TeV/fm}^{3} \nonumber \\
	&& \Longrightarrow\, B^{1/4} \sim 10\,{\rm GeV}/(\hbar c)^{3/4}.
\end{eqnarray}
The very high density contribution from the bag constant $4B/c^2 \sim 10^{5}$~TeV$\,c^{-2}fm^{-3}\sim10^{23}$~g/cm$^3$
might seem a bit peculiar. But then it should be kept in mind that the density contribution from the bag constant
in deconfined quark matter is $\sim 10^{15}$~g/cm$^3$, which is a large fraction of the maximum density in any type of
star composed of deconfined quark matter. So the high density is not that peculiar. On the contrary, if something is
to be expected, it should be that $B$ is much higher for preon matter than for quark matter, since the typical
``preon bag" is much smaller and more dense than a hadron. In the following, for simplicity, I put $\hbar c=1$ for the
bag constant and express $B^{1/4}$ in eV.

The density introduced by the bag constant is of the same order of magnitude as the density used in the mass-radius
estimate above. The improvement here is the transition to a proper EOS for fermions; the possibility to apply the EOS
in a general relativistic framework, for the analysis of mass-radius relations and stability. In addition to the full
general relativistic analysis, the mass and radius can be estimated from first principles as a function of the bag
constant~\cite{Banerjee:00}. The result is somewhat similar to the original Chandrasekhar limit, but the role of the
fermion mass is replaced by the bag constant $B$,
\begin{eqnarray}
M &=& \frac{16 \pi}{3 c^2} B R^3, \label{mchandra} \\
R &=& \frac{3 c^2}{16 \sqrt{\pi G B}}. \label{rchandra}
\end{eqnarray}
Inserting $B^{1/4} \sim 10$~GeV in~\eqns{mchandra}{rchandra}, an estimate for the (maximum) mass
$M \simeq 10^2$~M$_\oplus$ and radius $R \sim 1$~m is obtained. This is consistent with the somewhat simpler
mass-radius estimate given above.

\begin{figure}[ht]
\epsfig{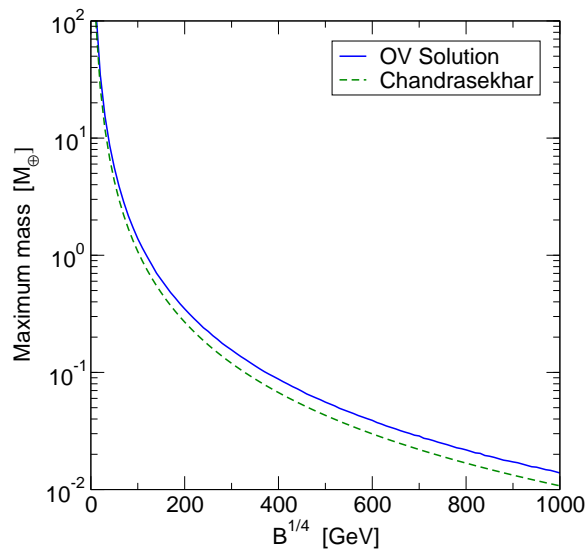}
\caption{The maximum mass of preon stars vs. the bag constant~$B$. The solid line represents the general relativistic
	OV solutions, while the dashed line represents the Newtonian (Chandrasekhar) estimate. Despite the high central
	density, the mass of these objects is below the Schwarzschild limit, as is always the case for static solutions
	to the stellar equations. M$_\oplus\simeq6\times10^{24}$~kg is the Earth mass.}
\label{maxmass}
\end{figure}

Since $B^{1/4}\sim 10$~GeV is only an order of magnitude estimate for the lower limit, the bag constant is considered
as a free parameter of the model, constrained by a lower limit of $B^{1/4}=10$~GeV and an upper limit chosen as
$B^{1/4}=1$~TeV. The latter value corresponds to a spatial extension of the electron of the order
$\sim\hbar c/10^3\,$TeV$\,\sim 10^{-22}$~m. In~\figs{maxmass}{maxradius} the maximum mass and radius of a preon star
are plotted vs. the bag constant.
%
%

\begin{figure}[ht]
\epsfig{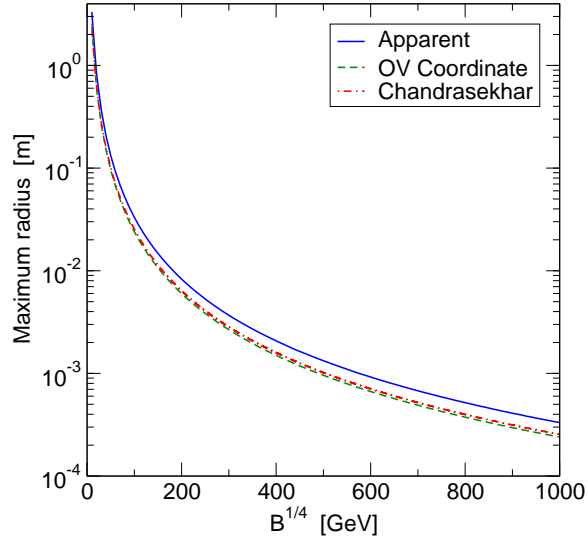}
\caption{The maximum radius of preon stars vs. the bag constant. The solid line is the
	``apparent" radius, $R^\infty = R / \sqrt{1-2 G M / Rc^2}$, as seen by a distant observer. The dashed line
	represents the general relativistic coordinate radius obtained from the OV solution. The dotted line
	represents the Newtonian (Chandrasekhar) estimate.}
\label{maxradius}
\end{figure}

A necessary, but not sufficient, condition for stability of a compact star, is that the total mass is an
increasing function of the central density, ${dM/d\rho_c>0}$. This condition is a consequence of
a generic microscopic relation known as Le Chatelier's principle. Roughly, this condition implies that
a slight compression or expansion of a star will result in a less favourable state, with higher total
energy. Obviously, this is a necessary condition for a stable equilibrium configuration. Equally important,
a star must be stable when subject to (small) radial oscillations, in the sense that the amplitude of the
oscillations must not grow spontaneously with time. Otherwise a small perturbation would bring about a
collapse of the star.

The equation for the analysis of such radial modes of oscillation is due to Chandrasekhar~\cite{Chandra:64}.
An overview of the theory, and some applications, can be found in~\cite{Misner:73}. A catalogue of various numerical
methods for solving the original set of equations can be found in~\cite{Bardeen:66}. However, a far more practical
form of the oscillation equations has been derived by Gondek {\it et al.}~\cite{Gondek:97}. The details of the
stability analysis can be found in~\cite{Hansson:04}. Here I summarise only the main points.

\begin{figure}[ht]
\epsfig{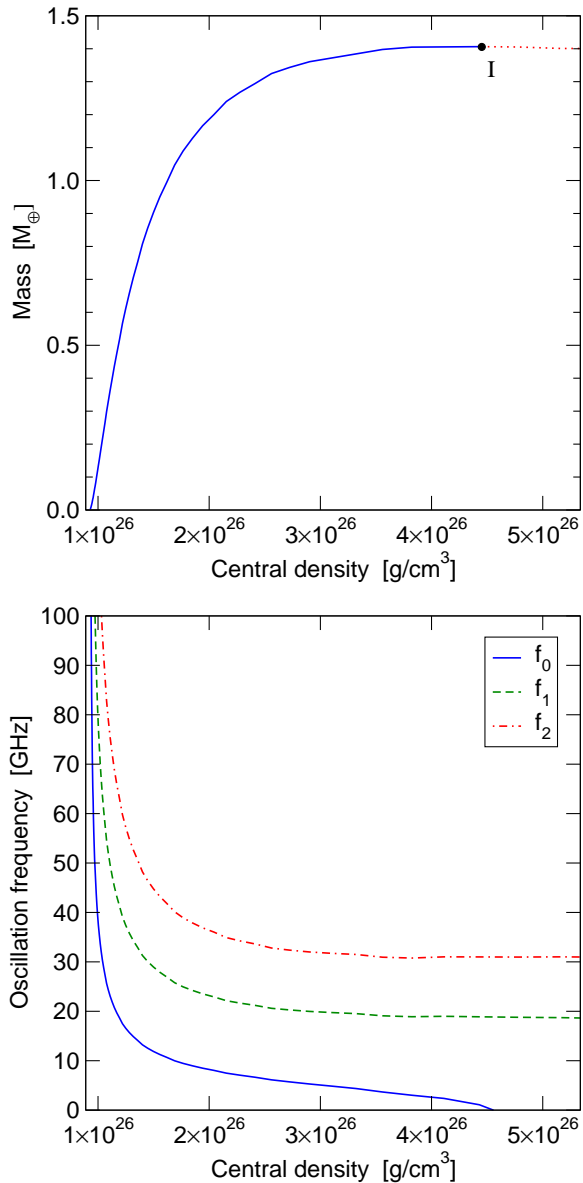}
\caption{The mass and the first three eigenmode oscillation frequencies ($f_0$,~$f_1$,~$f_2$) vs. the central
	density $\rho_c$ of preon stars. Here, a fixed value of $B^{1/4}=100$~GeV has been used. For the maximum mass
	configuration (I) the fundamental (nodeless) mode, $f_0$, has zero frequency, indicating the onset of instability.
	Preon stars with mass below the maximum of this sequence are stable.}
\label{b100gev}
\end{figure}

Assuming a time dependence of the radial displacement of fluid elements of the form $e^{i\omega t}$, the equation
governing the radial oscillations is found to be a Sturm-Liouville eigenvalue equation for $\omega^2$. A necessary,
and sufficient condition for stability is that all $\omega_i^2$ are positive, since imaginary frequencies give
exponentially increasing amplitudes. Furthermore, since $\omega_i^2$ are eigenvalues of a Sturm-Liouville equation, it
turns out that it is sufficient to prove that the fundamental (nodeless) mode, $\omega^2_0$, is positive for a
star to be stable. In~\fig{b100gev}, the first three oscillation frequencies, $f_i=\omega_i/2\pi$, for various stellar
%
%
configurations with $B^{1/4}=100$~GeV are plotted. In agreement with the theorem of Wheeler {\it et al.}~\cite{Harrison:65}
the onset of instability is the point of maximum mass, as $\omega_0^2$ becomes negative for higher central densities.
Thus, for $B^{1/4}=100$~GeV, preon stars are stable up to the maximum mass configuration. The same is true for other
values of $B$~\cite{Hansson:04}.

\begin{figure}[ht]
\epsfig{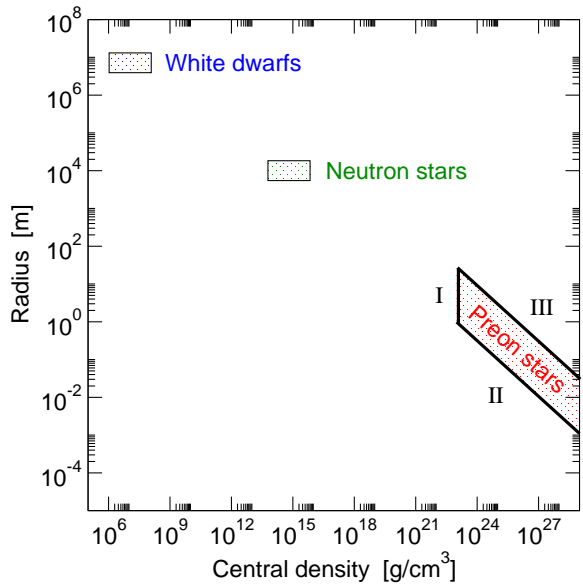}
\caption{The different types of compact stars traditionally considered in astrophysics are white dwarfs and 
	neutron stars (including quark and hybrid stars). In white dwarfs, electrons provide the Fermi pressure
	counterbalancing gravity. In neutron stars, the neutrons (quarks, hyperons etc.) play this role. If	quarks
	and leptons are composite particles, a new class of compact stars (preon stars) could exist. The minimum
	density (I) of preon stars is roughly given by the minimum density of leptons and quarks. The minimum size
	(II) for a given central density is due to the Schwarzschild radius (actually $4/3$ of it) and a maximum
	size (III) exists due to instability.}
\label{classes}
\end{figure}

Despite the large uncertainty regarding preon interactions, here manifested as a large uncertainty in the
bag constant, preon stars should have central densities beyond $10^{23}$~g/cm$^3$. This makes preon stars fundamentally
different from the traditional types of compact stars, since such high central densities implies that the stars
must be very small and light in order to be stable, see~\fig{classes}.
%
%


\section{Formation, detection and astrophysical consequences}

The list of possible connections between the properties of the fundamental particles and the large scale
structures in the universe is long. However, beyond a density of $\sim 10^{23}$~g/cm$^3$, not much has been
proposed, since there are strong arguments against the existence of stable objects beyond
$\rho \sim 10^{16}$~g/cm$^3$. That is, if quarks and leptons are fundamental entities.

If preons exist and objects composed of preon matter as small and light as suggested here are stable,
density fluctuations in the early universe might have produced primordial preon stars (or ``nuggets").
As this material did not take part in the ensuing nucleosynthesis, the abundance of preon nuggets is not
constrained by the hot big bang model bounds on baryonic matter. Also, preon nuggets are immune to Hawking
radiation~\cite{Hawking:75}, which rapidly evaporates small primordial black holes, making it possible for
preon nuggets to survive to our epoch. They can therefore serve as the mysterious dark matter needed in many
dynamical contexts in astrophysics and cosmology~\cite{Turner:00,Bergstrom:00}. The idea that preons could
be connected to dark matter is already recognized in the literature~\cite{Burdyuzha:98,Lalakulich:98}, but
the picture presented here is rather different.

The Friedmann equation for the early universe is $H^2(t)=8\pi G\rho/3$ and in the radiation-dominated epoch
$\rho c^2 \sim T^4$ (Boltzmann's law). When including the number of internal degrees of freedom $g_{eff}$, an
expression for the Hubble parameter $H$, in units where $\hbar=c=k_B=1$, is~\cite{Bergstrom:04}:
\begin{equation}
H \simeq 1.66 \sqrt{g_{eff}}\frac{T^2}{m_{pl}}.
\end{equation}
Here $T$ is the temperature in eV and $m_{pl} \simeq 1.2 \times 10^{19}$~GeV/$c^2$ the Planck mass. For the SM,
the fermions and the gauge and Higgs bosons give $g_{eff}(T=1\,{\rm TeV}) = 106.75$. In the preon phase,
this number should be smaller, say $g_{eff} \sim 10$ for simplicity. Then the Hubble radius at a temperature
of $1$~TeV is $H^{-1} \sim 1$~mm and the mass within the Horizon (a causally connected region) is
$\rho H^{-3} \sim 10^{-1}$~M$_\oplus$. This is the maximum mass of any structure that could have formed in
this early epoch. Hence, the maximum mass within causally connected regions, at the minimum temperature when deconfined
preon matter might have formed preon nuggets (and the particles of the SM), is of the correct order of magnitude
for stable configurations.

A potential problem is that the Jeans length, which defines the minimum length scale of regions that can
contract gravitationally, was roughly of the same order of magnitude as the Hubble radius at that temperature.
The Jeans length $\lambda_J$ is~\cite{Bergstrom:04}:
\begin{equation}
\lambda_J = v_s\sqrt{\frac{\pi}{G\rho_0}},
\end{equation}
where $v_s$ is the speed of sound and $\rho_0$ the average background density. For a relativistic fluid
with EOS $\rho c^2 = 3p + 4B$, the speed of sound is $v_s=c/\sqrt{3}$ and $\lambda_J \sim 1$~mm~$\sim H^{-1}$.
However, considering the high level of approximation used here, this is not yet a serious problem. It merely shows
that the numbers are in the correct intervals.

But, perhaps it will be the other way around. After all, Popper's idea that we make progress by falsifying
theories is not always true. By utilizing gamma-ray bursts (GRB) or white dwarfs in the large magellanic cloud
as light sources, gravitational lenses with very small masses produce a diffraction signal in the spectrum,
which might be observable~\cite{Gould:92,Stanek:93,Ulmer:95,Nemiroff:95}. For a lens with mass within the range
$10^{-16}$~M$_\odot \leq M \leq 10^{-11}$~M$_\odot$, the angular separation of images would be in the
femto-arcsec range (femtolensing), and for more massive lenses, $M \leq 10^{-7}$~M$_\odot$, the angular separation
is in the pico-arcsec range (picolensing). The mass within the Hubble radius at $T=1$~TeV is
$\sim 10^{-1}$~M$_\oplus\sim 10^{-7}$~M$_\odot$. This roughly defines the maximum mass of preon nuggets that could
be abundant enough to be observed as gravitational lenses. Hence, preon nuggets fall in the correct mass range for
picolensing and femtolensing.

\begin{figure}[ht]
\epsfig{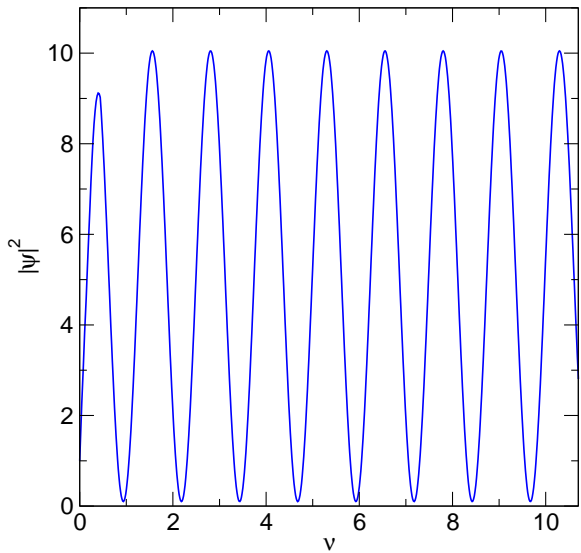}
\caption{$|\psi|^2$, the magnification of a distant light source vs. the dimensionless frequency, $\nu = \tilde{\nu}(1+z_L)2GM/c^3$,
	due to gravitational lensing by an intermediate preon star. The flux entering the detector is obtained by
	multiplying the magnification, $|\psi|^2$, with the flux in the absence of a lens. For a $10^{-6}$~M$_\oplus$ preon star
	located in the halo of our galaxy, $\nu=1$ corresponds to a photon energy of $0.14$~keV. See the text for
	further details.}
\label{magnification}
\end{figure}

In~\fig{magnification} the magnification of a distant light source due to gravitational lensing by an intermediate
preon star is plotted as a function of the dimensionless frequency:
\begin{equation}
\nu = \frac{\tilde{\nu}(1+z_L)2GM}{c^3}.
\end{equation}
Here, $M$ is the mass of the lens, $z_L$ the redshift of (distance to) the lens and $\tilde{\nu}$ the frequency
of light. This result was calculated with a physical-optics model, as described in~\cite{Ulmer:95}. In principle,
the time dependent amplitude due to a single light pulse from the source was calculated, and then the power
spectrum was obtained by a Fourier transform of the amplitude. The magnification is normalized to a unit flux
in the absence of a lens, {\it i.e.}, the signal is the observed flux of the source times the magnification. The
exact shape of the magnification curve depends on the relative position of the source and lens. Here the source is
slightly off-axis (corresponding to $\theta=0.2$ in~\cite{Ulmer:95}).
%
%

As mentioned in~\cite{Hansson:04}, preon stars could also form in the collapse of normal massive stars,
if the collapse is slightly too powerful for the core to stabilize as a neutron star, but not sufficiently
violent for the formation of a black hole. Due to the potentially very large magnetic field and rapid rotation
of preon stars formed in this way, the astrophysical consequences could be important, {\it e.g.}, for
acceleration of ultra-high energy (UHE) cosmic rays. However, even though the idea might be appealing at first
sight, the possibility to expel such a large fraction of the mass of the progenitor star needs to be better
understood. What should be noted here is merely a potential connection to UHE cosmic rays, which might provide
a second means for locating and observing preon stars.


\section{Conclusions}

If there is a deeper layer of fermionic constituents (preons), below that of quarks and leptons, a new class
of stable compact stars could exist. By fitting a simple equation of state for fermions to the minimum energy
density of an electron, the maximum mass for stars composed of preons can be estimated to $\sim 10^{2}$~Earth
masses and the maximum radius to $\sim 1$~m. The minimum central density is of the order of
$\sim 10^{23}$~g/cm$^3$. Preon stars with a maximum mass of $\sim 10^{-1}$~Earth masses
and radius $\sim 1$~mm could have been formed by the primordial density fluctuations in the early universe. By
utilizing gamma-ray bursts, or white dwarfs in the large magellanic cloud as light sources, an intermediate preon
star would produce a diffraction signal in the spectrum, which might be observable. An observation of an
object as dense as a preon star would be a direct evidence for a new state of matter, which is not composed of
quarks and leptons. Due to the need for observational clues in the cold dark matter sector, this could prove
compositeness plausible, without much dedicated effort. This approach might complement direct tests of
preon models~\cite{PreonTrinity:02} performed at particle accelerators.


\section{Acknowledgements}

I acknowledge support from the Swedish National Graduate School of Space Technology. I thank S.~Fredriksson
and J.~Hansson for several useful discussions and for reading the manuscript. I thank M.~Alford for providing
me with the data tables of the high density nuclear matter equations of state, J.~Goodman for useful discussions
regarding the femtolensing signature, and J.~Bourjaily, D.~Casadei, A.~Geiser, A.~Giazotto, and J.~Kamenik for
interesting discussions about the astrophysical consequences of preon stars. Finally, I thank G.~'t~Hooft and
A.~Zichichi for organizing an excellent 42nd course of the international school of subnuclear physics, and for
the award that was designated this work.




\begin{thebibliography}{5}


\bibitem{Weber:04} F. Weber, {\it Strange Quark Matter and Compact Stars},
	astro-ph/0407155 (submitted).


\bibitem{Harrison:65} B.K. Harrison, K.S. Thorne, M. Wakano \& J.A. Wheeler,
	{\it Gravitation Theory and Gravitational Collapse} (University of Chicago Press, Chicago, 1965).


\bibitem{Gross:73} D.J. Gross \& F. Wilczek, Phys. Rev. Lett. {\bf 30}, 1323 (1973).

\bibitem{Politzer:73} H.D. Politzer, Phys. Rev. Lett. {\bf 30}, 1346 (1973).


\bibitem{Gerlach:68} U.H. Gerlach, Phys. Rev. {\bf 172}, 1325 (1968), \\
	U.H. Gerlach, Ph.D. thesis, Princeton University, 1968.

\bibitem{Glendenning:00} N.K. Glendenning \& C. Kettner, Astron. Astrophys. {\bf 353}, L9 (2000).

\bibitem{Schertler:00} K. Schertler {\it et al.}, Nucl. Phys. {\bf A 677}, 463 (2000), astro-ph/0001467.


\bibitem{Hansson:04} J. Hansson \& F. Sandin, {\it Preon stars: a new class of cosmic compact objects}, 2004, astro-ph/0410417.


\bibitem{Fredriksson:03} S. Fredriksson, in {\it Proc. of the Fourth Tegernsee Int. Conf.
	on Particle Physics Beyond the Standard Model, 2004}, ed. by H.-V. Klapdor-Kleingrothaus
	(Springer-Verlag, Heidelberg, 2004), p. 211, hep-ph/0309213.

\bibitem{dSouza:92} I.A. D'Souza \& C.S. Kalman, {\it Preons} (World Scientific, Singapore, 1992).

\bibitem{tHooft:79} G. 't Hooft, Carg\`{e}se Lecture Notes, 1979.

\bibitem{Barbieri:80} R. Barbieri, L. Maiani \& R. Petronzio, Phys. Lett. {\bf B 96}, 63 (1980).

\bibitem{PreonTrinity:02} J.-J. Dugne, S. Fredriksson \& J. Hansson, Europhys. Lett. {\bf 57}, 188 (2002).


\bibitem{OV} J.R. Oppenheimer \& G. Volkoff, Phys. Rev. {\bf 55}, 374 (1939).


\bibitem{NV:71} J.W. Negele \& D. Vautherin, Nucl. Phys. {\bf A 207}, 298 (1973).

\bibitem{APR:98} A. Akmal, V.R. Pandharipande \& D.G. Ravenhall, Phys. Rev. {\bf C 58}, 1804 (1998), nucl-th/9804027.


\bibitem{Kettner:95} C. Kettner {\it et al.}, Phys. Rev. {\bf D 51}, 1440 (1995).

\bibitem{Prisznyak:94} M. Prisznyak, B. Lukacs \& P. Levai, {\it Are There Top Quarks in Superdense Hybrid Stars?}, astro-ph/9412052.


\bibitem{Chodos:74} A. Chodos {\it et al.}, Phys. Rev. {\bf D 9}, 3471 (1974).


\bibitem{Glendenning:97} N.K. Glendenning, {\it Compact Stars} (Springer-Verlag, New York, 1997).


\bibitem{Banerjee:00} S. Banerjee, S.K. Ghosh \& S. Raha, J. Phys. {\bf G 26}, L1 (2000), astro-ph/0001246.


\bibitem{Chandra:64} S. Chandrasekhar, Phys. Rev. Lett. {\bf 12}, 114 (1964).

\bibitem{Misner:73} C.W. Misner, K.S. Thorne \& J.A. Wheeler, {\it Gravitation} (Freeman and Co., San Francisco, 1973).

\bibitem{Bardeen:66} J.M. Bardeen, K.S. Thorne \& D.W. Meltzer, Astrophys. J. {\bf 145}, 505 (1966).

\bibitem{Gondek:97} D. Gondek, P. Haensel \& J.L. Zdunik, Astron. Astrophys. {\bf 325}, 217 (1997), astro-ph/9705157.


\bibitem{Hawking:75} S.W. Hawking, Commun. Math. Phys. {\bf 43}, 199 (1975).


\bibitem{Turner:00} M.S. Turner, Phys. Rep. {\bf 333}, 619 (2000).

\bibitem{Bergstrom:00} L. Bergstr\"{o}m, Rep. Progr. Phys. {\bf 63}, 793 (2000).

\bibitem{Burdyuzha:98} V. Burdyuzha {\it et al.}, in {\it Proc. of the Second Int. Workshop on
	Particle Physics and the Early Universe, 1999}, ed. by D.O. Caldwell (Springer-Verlag, 1999),
	p. 392, astro-ph/9912555.

\bibitem{Lalakulich:98} O. Lalakulich \& G. Vereshkov, in {\it Proc. of the Second Int. Conf.
	on Dark Matter in Astro and Particle Physics, 1999}, ed. by  H.-V. Klapdor-Kleingrothaus
	and L. Baudis (IOP Publishing, Great Yarmouth, 1999), p. 668.
 

\bibitem{Bergstrom:04} L. Bergstr\"{o}m \& A. Goobar, {\it Cosmology and Particle Astrophysics}, 2nd ed.
	(Springer-Verlag, Germany, 2004).
	

\bibitem{Gould:92} A. Gould, ApJ {\bf 386}, L5 (1992).

\bibitem{Stanek:93} K.Z. Stanek, B. Paczy\'{n}ski \& J. Goodman, ApJ {\bf 413}, L7 (1993).

\bibitem{Ulmer:95} A. Ulmer \& J. Goodman, ApJ {\bf 442}, 67 (1995), astro-ph/9406042.


\bibitem{Nemiroff:95} R.J. Nemiroff \& A. Gould, {\it Probing for MACHOs of Mass
	$10^{-15}$M$_\odot - 10^{-7}$M$_\odot$ with Gamma-Ray Burst Parallax Spacecraft},
	astro-ph/9505019.

\end{thebibliography}
\end{document}